\documentclass[12pt,a4paper]{article}
\usepackage{amsmath}
\usepackage{graphicx}
\usepackage{times}
\begin{document}
\begin{titlepage}
\title{Generalized upper bound for  inelastic diffraction}
\author{ S.M. Troshin and N.E. Tyurin\\[1ex]
\small  \it SRC IHEP of NRC ``Kurchatov Institute''\\
\small  \it Protvino, 142281, Russian Federation}
\normalsize
\date{}
\maketitle

\begin{abstract}
For  the inelastic diffraction, we  obtain an upper bound valid in the whole range of the elastic scattering amplitude variation allowed by unitarity. We discuss  the energy dependence of the inelastic diffractive cross-section on the base of this bound and recent LHC data. 
\end{abstract}
\end{titlepage}
\setcounter{page}{2}
\section*{Introduction}
The recent experimental measurements of the global observables at the LHC in proton--proton collisions have 
confirmed the trends observed at lower energies,
namely, continuous increase of the total, elastic and inelastic cross--sections in the new energy region. 
Those experiments have brought us  closer to clarification of an elusive asymptotic regime of 
strong interactions at fixed, low values of the momentum transferred  and increasing energy of collision. 
This regime belongs to the soft interaction region where perturbative QCD does not work. Nonperturbative QCD is still not able to provide calculations of these global observables despite that lattice QCD has successfully predicted various static characteristics of hadrons\footnote{We are grateful to R.N. Rogalev for discussion on the present status and problems in the lattice QCD calculations of soft hadron interactions.}. 

However, all hadron interactions  are under constraints which follow
from the general principles of theory such as unitarity and analyticity. Those principles are valid for any quantum field theory including QCD.
The known for a long time properties of analyticity and unitarity of the scattering matrix 
could lead to assumption that strong interactions are ``as strong as possible'' \cite{chew, chew1} and therefore the 
Froissart-Martin bound \cite{froi,martin} for the total cross-sections 
should be saturated at asymptotical energies. 
Then, the explicit functional energy behavior of the  total cross-sections is often taken to follow $\ln^2 s$-dependence 
but 
the value of the factor in front of $\ln^2 s$ is still a remaining problem. 
The magnitude of this factor is an important issue since it is
correlated with a particular choice  for the upper limit for the elastic partial amplitude. Namely,  this limit (it is $1/2$ in the case of the pure imaginary elastic scattering amplitude) may correspond to the
maximum of the inelastic channel contribution to the elastic unitarity and leads to the ratio
\begin{equation}\label{bd}
  \sigma_{el}(s)/\sigma_{tot}(s)\to 1/2,
\end{equation}
or it can correspond to a maximal value of the partial amplitude allowed by unitarity (i.e. unity) resulting in the asymptotical limit
\begin{equation}\label{rd}
  \sigma_{el}(s)/\sigma_{tot}(s)\to 1.
\end{equation}
In the above equations $\sigma_{el}(s)$ denotes integral cross--section of the elastic scattering and $\sigma_{tot}(s)=\sigma_{el}(s)+
\sigma_{inel}(s)$, where $\sigma_{inel}(s)$ is the total cross--section of inelastic interactions. Note, that Eq. (\ref{rd}) does not preclude asymptotical growth of $\sigma_{inel}(s)$, it just means that the inelastic cross-sections growth
$\sigma_{inel}(s) \sim \ln s$ is slower than the growth of the elastic cross-section, i.e. $\sigma_{el}(s)\sim \ln^2 s$. The experimental values for total cross-section $\sigma_{tot}(s)$ are obtained from the differential cross-section of elastic scattering extrapolated to the point $t=0$ through optical theorem.  

In the impact parameter representation Eq. (\ref{bd}) corresponds to the limiting case of the so called BEL picture. The acronym BEL has the meaning that the interaction region of
protons becomes Blacker, Edgier and Larger with increasing energy.  This kind of energy evolution was proposed and discussed in \cite{henzi}.
Such  behaviour corresponds to  the presupposed absorptive nature of the scattering when the elastic scattering partial amplitude would never exceed the black disc limit of $1/2$.
With this assumption of  the absorptive scattering domination the original Froissart-Martin bound for the total cross-sections 
has been improved by factor of $1/2$.  An origin for appearance of this factor is an upper bound for the total inelastic cross-section reduced by factor of 4 (cf. \cite{mart}).
Thus, the related ratio of the cross--sections $ \sigma_{el}(s)/\sigma_{tot}(s)$ is an important function of energy, e.g. it stands in front of $\ln^2 s$ in the asymptotical bound on the total 
cross-section \cite{roy,roy1}:
\begin{equation}\label{sing}
\sigma_{tot}(s)\leq \frac{4\pi}{t_0}\left(\frac {\sigma_{el}(s)}{\sigma_{tot}(s)}\right)\left[\ln\left(\frac{s}{\sigma_{el}(s)}\right)\right]^2
\left[1+\left(\frac{\mbox{Re} F(s, t=0)}{\mbox{Im} F(s, t=0)}\right)^2\right]^{-1},
\end{equation}
where $F(s,t)$ is the amplitude of the elastic scattering process $pp\to pp$. The elastic scattering amplitude $F(s,t)$ is related to the amplitude in the impact parameter representation $f(s,b)$ through the  Fourier-Bessel transformation. 

Since our consideration is a qualitative one, we will adopt for simplicity  that the scale of $s$ in the above bound is  determined by the energy-independent value $s_0=1$ GeV$^2$ (i.e. $s/s_0$ is replaced by $s$ and the latter is considered to be dimensionless, despite that this scale is an energy-dependent one and is determined by $ \sigma_{el}(s)$) in the above bound for the total cross--section. In Eq. (\ref{sing})   
$\sqrt{t_0}$  is the mass of the lowest state in the $t$ channel in this bound\footnote{For most processes, the factor in Eq. (\ref{sing}) is determined by a pion mass, i.e.
${t_0}=4m^2_{\pi}$. Assumption of saturation of this bound leads to necessity of a significant shift of $t_0$ into the region of larger values, and it corresponds to replacement of the pion mass by a mass of  a heavy particle  in order to fit the experimental data.}. 

In this note, an upper bound for the cross--section of inelastic diffraction is discussed. We point out a problem with an existing upper bound for the cross--section of inelastic diffraction. In fact, the existing, Pumplin bound  \cite{pumplin}, has been derived under assumption of the absorptive nature of diffraction at all the energies. A generalized bound which is free  from such restriction and valid in the whole range of the elastic scattering amplitude variation allowed by unitarity is obtained. We also discuss its implications on the model ground.
\section{The absorptive and reflective scattering modes}
A distinctive feature of the impact parameter representation  is a diagonal form of the unitarity equation  for the elastic scattering amplitude $f(s,b)$, i.e.  
\begin{equation}\label{unit}
\mbox{Im} f(s,b)=|f(s,b)|^{2}+h_{inel}(s,b).
\end{equation}
Eq. (\ref{unit}) is valid at high energies with ${\cal O}(1/s)$ precision \cite{gold}. 
The $|f(s,b)|^{2}$ is the elastic channel contribution, while the inelastic overlap function $h_{inel}(s,b)$ 
includes  all contributions from   the intermediate inelastic channels into unitarity relation.
The elastic scattering $S$-matrix element in the impact parameter representation is  related to the elastic scattering amplitude $f(s,b)$ by the equation $S(s,b)=1+2if(s,b)$\footnote{This relation determines normalization of the elastic scattering amplitude $f(s,b)$} and it
can be represented in the form
\[S(s,b)=\kappa(s,b)\exp[2i\delta(s,b)]\]
which includes the two real functions $\kappa(s,b)$ and $\delta(s,b)$. 
The function $\kappa$ (it varies in the region $0\leq \kappa \leq 1$) is called an absorption factor: its value $\kappa=0$ corresponds to a complete absorption of the initial state. 

As it was already mentioned,  we assume the pure imaginary scattering amplitude\footnote{It should be noted that saturation of the black--disc limit or the unitarity limit leads to a vanishing real part of the scattering amplitude, $\mbox{Re} f\to 0$ in the region where the both limits are saturated \cite{tr}. The recent data \cite{tote} on the precise measurements of the ratio of the real to the imaginary part of the forward  amplitude are consistent with decreasing energy dependence of this ratio.} and perform the replacement $f\to if$ at high energies. 
We should also mention that the Pumplin bound discussed below
has been derived with the approximation of  pure imaginary amplitudes of the elastic and diffractive scattering. The elastic scattering amplitude $f(s,b)$ is related then to the elastic scattering matrix element by the relation $S(s,b)=1-2f(s,b)$, i.e.
the function $S(s,b)$ is real, but it should not have a definite sign in the whole range of the amplitude variation allowed by unitarity.
On the base of the  existing experimental trends a monotonic, without oscillations over $s$, increase of the elastic scattering amplitude $f(s,b)$ with the energy is assumed. 

In fact, the choice of the particular elastic scattering mode, namely, absorptive  or reflective one \cite{intja}, depends on the sign of the function $S(s,b)$, in another words, on the value of the phase $\delta(s,b)$
\cite{ttprd}. The widely used picture is based on the limit  $S(s,b)\to 0$  at  fixed impact parameter $b$ and $s\to \infty$. This  is known as a black-disc limit. The elastic scattering  is a completely absorptive one in this mode and the function $S(s,b)$ is
always non-negative and  the limitation $f(s,b)  \leq 1/2$ is implied.   

There is  another mode (reflective scattering) when the function $S$ becomes negative with energy growth. It has the limiting behavior $ S(s,b)\to -1$ at fixed $b$  and $s\to \infty$, i.e.  $\kappa \to 1$ and $\delta = \pi/2$ and the amplitude  varies in the range  $1/2<f(s,b)  \leq 1$ since the function $S$ is negative. The non-zero phase $\delta = \pi/2$ can be interpreted as the geometric phase and its appearance is related to the presence of a dynamical singularity \cite{intja,arh13}.

Which mode will be realized at the asymptotical energies is unclear now. The claims on the equipartition of the elastic and inelastic collision probabilities   at $s\to \infty$  are the model-dependent ones. Moreover, conclusions on such equipartition  are ambiguous without impact parameter analysis performed. The  respective conclusions \cite{blok} are based on the forward scattering data only and are not trustworthy therefore. 
The base for such statement is grounded on the fact that
the forward scattering observables are represented by definite integrals over impact parameter. It is obvious,  that an integration does not allow one  to reconstruct an integrand since different functions can lead to the same results after integration  (cf. e.g. \cite{white} where the equipartition $\sigma_{el}(s)/\sigma_{tot}(s)=\sigma_{inel}(s)/\sigma_{tot}(s)=1/2$ at 
$s \to \infty$ has been obtained just for a gaussian exponential dependence of the profile function saturating unitarity limit. Such function has nothing to do with the black disc model).  

\section{Sub-leading role of inelastic diffraction  in the absorptive mode} 
It is known that soft hadron interaction dynamics  is determined by the nonperturbative QCD. Since it does not provide calculation recipe at the moment, the essential role under discussion of soft processes belongs to general principles of the theory such as unitarity and analyticity, as it was mentioned in the Introduction.  We start with discussion of an upper bound
in the impact parameter representation. 
The assumption on absorptive scattering domination at all the energies including asymptotics  was an essential point
under  derivation of the Pumplin bound \cite{pumplin}
for the inelastic diffraction:
\begin{equation}\label{pump}
 \sigma_{diff}(s,b)\leq \frac{1}{2}\sigma_{tot}(s,b)-\sigma_{el}(s,b),
\end{equation}
where  \[\sigma_{diff}(s,b)\equiv \frac{1}{4\pi}\frac{d\sigma_{diff}}{db^2}\] is the total cross--section of  all the inelastic diffractive processes in the impact parameter 
representation and \[ \sigma_{tot}(s,b)\equiv \frac{1}{4\pi}\frac{d\sigma_{tot}}{db^2}\,\, ,\,\,  \sigma_{el}(s,b)\equiv \frac{1}{4\pi}\frac{d\sigma_{el}}{db^2}.\]
The bound Eq. (\ref{pump}) was obtained in the framework of the Good--Walker formalism for the inelastic diffraction \cite{gwalk} and is  based on the  eikonal form of the diffractive amplitudes. This form can be considered as  one of the realizations of the absorptive scattering mode. 
Eq. (\ref{pump})  is valid for  each value of the impact parameter of the collision $b$, i.e. it is diagonal in $b$-space and integration over impact parameter provides:
\begin{equation}\label{pumpint}
 \sigma_{diff}(s)\leq \frac{1}{2}\sigma_{tot}(s)-\sigma_{el}(s).
\end{equation}
It should be emphasized again  that eqs. (\ref{bd}) and (\ref{pumpint}) are to be fulfilled
if the hadron scattering picture corresponding to the black-disc limit is valid at asymptotical energies, i.e. 
\begin{equation}\label{bdin}
  \sigma_{inel}(s)/\sigma_{tot}(s)\to 1/2
\end{equation}
while
\begin{equation}\label{bdind}
  \sigma_{diff}(s)/\sigma_{tot}(s)\to 0
\end{equation}
and
\begin{equation}\label{bdindi}
  \sigma_{diff}(s)/\sigma_{inel}(s)\to 0
  \end{equation}
at $s \to \infty$.
 It is difficult to reconcile those  limits. Indeed,  $\sigma_{diff}(s)$ is, by definition\footnote{Generally, any type of inelastic diffraction is  associated with one or several  Pomeron exchanges. Cf. for discussion \cite{pred}.}, a leading part of the inelastic
cross--section $\sigma_{inel}(s)$ and the LHC experimental data demonstrate approximate
energy--independence of ratio
$ \sigma_{diff}(s)/\sigma_{inel}(s)$   \cite{alice, lipari}.

In contrast to its definition and the available data, one should conclude then, that the inelastic diffraction corresponds, in fact, to  a sub-leading mechanism 
 in the inelastic cross-section and the leading role in the growth of $\sigma_{inel}(s)$
 belongs to the  nondiffractive inelastic processes. 
 Such a statement is not easy to reconcile with the definition of the diffractive processes and their energy dependence, i.e. the  eq. (\ref{bdindi}) is not in  favor of the black-disc limit saturation by the partial scattering amplitude at $s\to\infty$.

We also note that   Eq. (\ref{pump}) can be rewritten in terms of $S(s,b)$  in the form
\begin{equation}\label{pbsf}
 \sigma_{diff}(s,b)\leq \frac{1}{4}S(s,b)(1-S(s,b)) 
 \end{equation}
and this inequality indicates that the Pumplin bound on  $\sigma_{diff}(s,b)$ cannot be applied in the region where $S(s,b)$ becomes negative.
\section{Generalized upper bound for  inelastic diffraction}
Apparently, there is no embarrassment  in the approach which  allows saturation of the unitarity limit. The limiting dependence Eq. (\ref{rd}) assumes an alternative option which corresponds to saturation of the unitarity limit for the elastic partial amplitude and can be interpreted as a reflective scattering \cite{intja}. Saturation of  unitarity can be associated with the strong coherent parton interactions in QCD relevant to confinement dynamics.   

The inelastic overlap function at the asymptotical energies will acquire a peripheral form in  the impact parameter representation  \cite{edn}.  This 
peripherality was treated as a manifestation of an emerging transparency in the central hadron collisions (or in vicinity of the  impact parameter  $b=0$ of the colliding particles) at very high energies. Later on, this interpretation has been generalized and specified in papers \cite{bd1,bd2,bd3,degr} where such a phenomenon  was related to antishadowing or reflection in the hadron interactions.  It should be noted that the  concept of the on-shell optical potential also lead up to conclusion on the central grayness in the inelastic overlap function  \cite{arriola}.
The peripheral form of the inelastic overlap function appears at high energies due to acquiring negative values by $S(s,b)$ with the collision energy increase:
\begin{equation}\label{der}
\frac{\partial h_{inel}(s,b)}{\partial b}=S(s,b)\frac{\partial f(s,b)}{\partial b} 
\end{equation}
Thus, the central profile of $f(s,b)$ transforms with the energy growth into a peripheral profile of the function $h_{inel}(s,b)$ due to negative values of $S(s,b)$. The appearing of peripheral form of the inelastic overlap function is an
 energy-dependent  effect. It happens at the values of energy $s>s_r$, where $s_r$ is solution of the equation
 \begin{equation}\label{sr}
S(s_r,b=0)=0
\end{equation} 
A recent analysis \cite{alkin} of the elastic scattering data obtained by the TOTEM Collaboration at  
$\sqrt{s}=7$ TeV \cite{totem} confirmed an existence  of this novel feature in strong interaction dynamics revealing that way transition   to  such scattering mode   (also referred as a resonant scattering \cite{anis}). 
A gradual transition to the REL picture,  acronym REL means that  the interaction region becomes Reflective (the term reflective means that the elastic scattering matrix element acquires negative values) close to the center ($b=0$) and simultaneously becomes  Edgier, Larger and completely black in the ring at periphery (at $b\neq 0$), seems to be observed by the TOTEM  under the measurements of the $d\sigma/dt$ in elastic $pp$--scattering \cite{totem}. The acronym REL was formed similar to the above acronym BEL. In this REL picture elastic partial amplitude exceeds black-disc limit at small values of $b$. Several phenomenological models are able to reproduce such transition and among them the one  based on the rational unitarization of the leading vacuum Regge--pole contribution with the intercept greater than unity 
\cite{edn} and similar models   known under the generic name of the unitarized supercritical Pomeron (cf. \cite{anis} for a recent  discussion and the references).  

The assumption that the unitarity limit instead of the black-disc limit is to be saturated asymptotically  leads to a relatively slower increase of the inelastic cross-section
\begin{equation}\label{bdind0}
  \sigma_{inel}(s)/\sigma_{tot}(s)\to 0
\end{equation}
 which allows one to keep considering inelastic diffraction as a leading mechanism  of the inelastic cross--sections growth.
It should be noted that the available experimental data are consistent with decreasing
dependence of the ratio $ \sigma_{inel}(s)/\sigma_{tot}(s)$ with energy.

The possibility that the  elastic amplitude exceeds the value of $1/2$ (which corresponds to the black-disc limit)   was discussed earlier in the framework of the rational 
unitarization on the base of the CDF data obtained at Tevatron \cite{bd2}. It should be noted that the value of $\mbox{Im} f(s,b=0)$ 
has increased from $0.36$  (CERN ISR) to $0.492\pm 0.008$ (Tevatron)  and it is just on the edge of the black-disc limit in the Tevatron energy domain\cite{girom}. 

The exceeding of the black-disc limit of $1/2$ turns the Pumplin bound to be groundless \cite{bd1}. But, this conclusion deserves to be more specified. In fact, the Pumplin bound is not valid only in  the particular region of the small and moderate values of the impact parameter  where the absorptive approach becomes not applicable. Such region of impact parameter values appears at very high energy.

Namely, the model-independent  reconstruction of the impact--parameter dependent quantities from the TOTEM  data demonstrates that the black-disc limit has been crossed in elastic scattering at small values of $b$ \cite{alkin}. In fact, the elastic scattering $S$-matrix 
element $S(s,b)$ is negative at $0<b<0.2$ fm and crosses the zero
at $b=0.2$ fm at $\sqrt{s}=7$ TeV. These estimates are consistent with the result of the Tevatron data analysis  \cite{girom}.  
It should be noted here
that the region of the negative values of $S(s,b)$ is determined by the interval $0<b<r(s)$. The function $r(s)$ is  the solution of the following equation:
\[
S(s,b=r(s))=0,
\]
 $r(s)=0$ at $s=s_r$.

The qualitative impact parameter dependencies of the functions $f(s,b)$ and $h_{inel}(s,b)$ at fixed energy $s_2>s_r$ are depicted in Figs. 1 and 2 and
the schematic energy evolution of the function $S(s,b)$
is depicted in Fig. 3 at three different energy values.
\begin{figure}[hbt]
\begin{center}
\resizebox{9cm}{!}{\includegraphics*{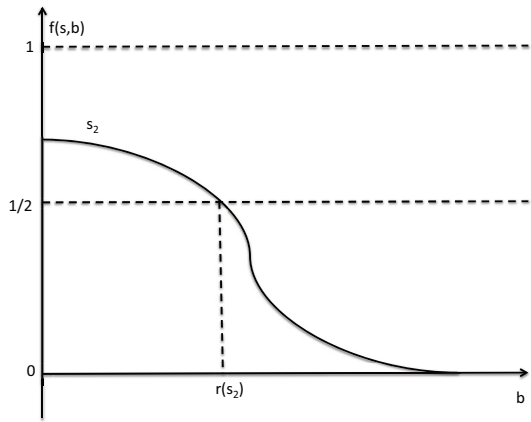}}
\end{center}
\vspace{-1cm}
\caption[ch1]{\small Qualitative impact-parameter dependence of $f(s,b)$ at $s=s_2$ ($s_2>s_r$).}
\end{figure}
\begin{figure}[hbt]
\begin{center}
\resizebox{9cm}{!}{\includegraphics*{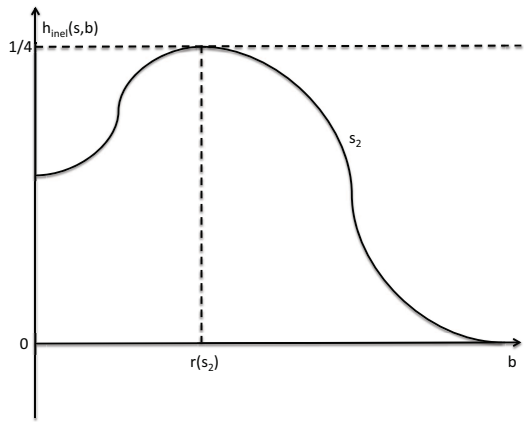}}
\end{center}
\vspace{-1cm}
\caption[ch1]{\small Qualitative impact-parameter dependence of $h_{inel}(s,b)$ at $s=s_2$ ($s_2>s_r$).}
\end{figure}
\begin{figure}[hbt]
\begin{center}
\resizebox{9cm}{!}{\includegraphics*{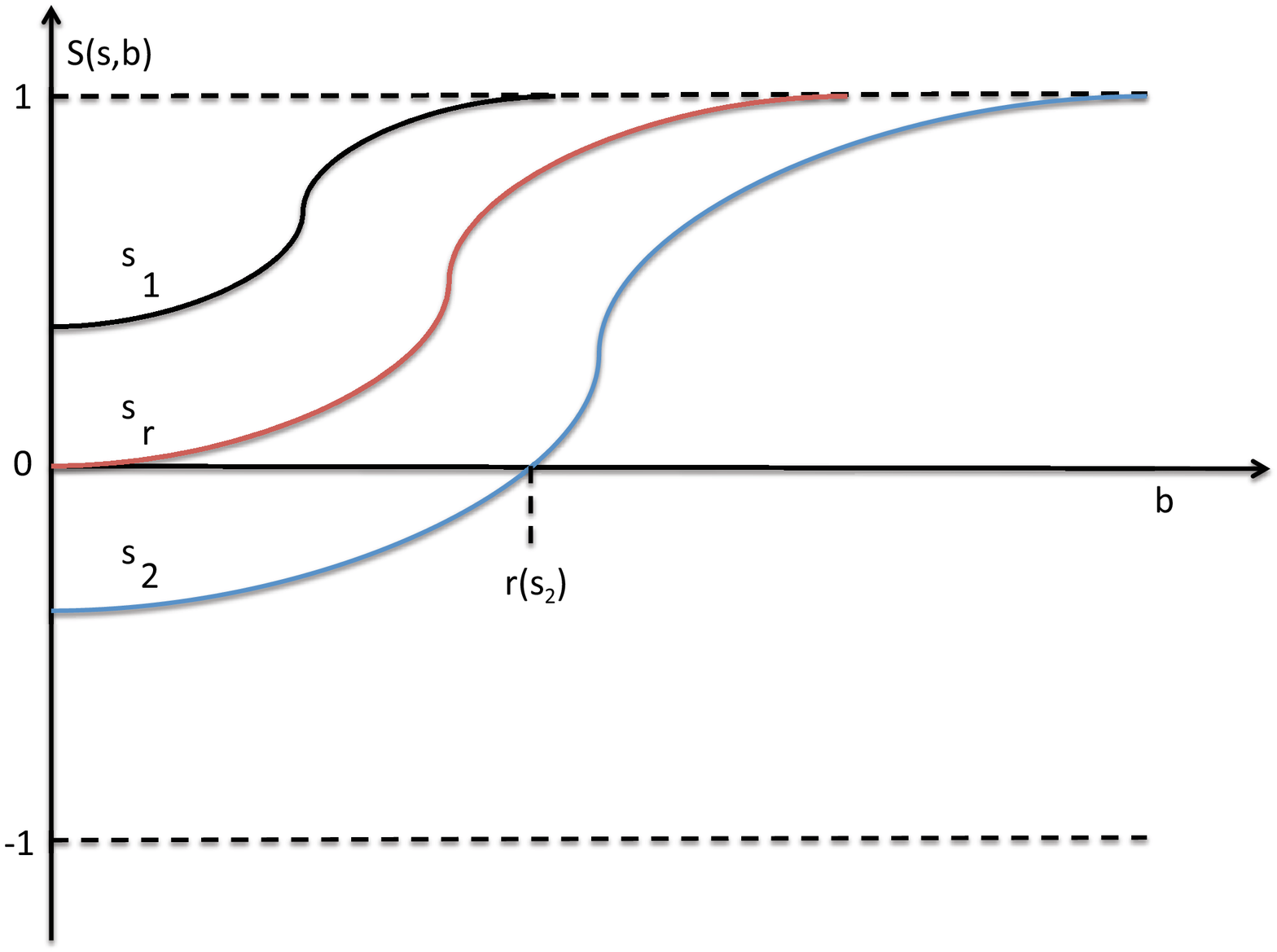}}
\end{center}
\vspace{-1cm}
\caption[ch1]{\small Schematic energy evolution of the impact-parameter dependence  $S(s,b)$.}
\end{figure}
In the impact parameter
range $0<b\leq r(s)$ only a trivial bound 
\begin{equation}\label{pbsb}
 \sigma_{diff}(s,b)\leq \sigma_{inel}(s,b)
\end{equation}
can be applied.
But, at $b\geq r(s)$ the scattering is absorptive and therefore the original Pumplin bound should be restored. 

However, the integrated bound is modified. Namely, in this case it should be written in the form
\begin{equation}\label{pumprefl}
\bar \sigma_{diff}(s)\leq \frac{1}{2}\bar \sigma_{tot}(s)-\bar \sigma_{el}(s),
 \end{equation}
where  $\bar \sigma_i(s)$ are the ``reduced'' cross-sections:
 \begin{equation}\label{genpb}
 \bar \sigma_i(s)\equiv \sigma_i(s) -8\pi\int_0^{r(s)}bdb\sigma_i(s,b),
 \end{equation}
 for $i\equiv diff, tot, el$, respectively. 
Combining Eqs. ( \ref{pbsb}) and  ( \ref{pumprefl}), the following inequalities relevant for the LHC energies, 
 can be easily  obtained:
 \begin{equation}\label{brfd}
 \sigma_{diff}(s) \leq \sigma_{inel}(s)- 2\pi\int_{r(s)}^\infty bdb[1-S(s,b)]
 \end{equation}
and
\begin{equation}\label{brfnd}
 \sigma_{ndiff}(s) \geq 2\pi\int_{r(s)}^\infty bdb[1-S(s,b)],
 \end{equation}
 where $\sigma_{ndiff}(s)$ is the total cross-section of the nondiffractive procsses, i.e. \[\sigma_{ndiff}(s)\equiv \sigma_{inel}(s)-\sigma_{diff}(s).\]
   The function $S(s,b)$ can be reconstructed from the experimental data
 on $d\sigma/dt$ in elastic $pp$-scattering. Using the TOTEM data at $\sqrt{s}=7$ TeV and  the value of $r(s)=0.2$ fm extracted from the analysis  \cite{alkin}, one obtains the magnitude of the upper bound on the value of $\sigma_{diff}(s)$ to be equal to $25.6$ mb at this energy.
The positive contribution of reflective scattering to the right hand side of Eq. (\ref{brfd}) at this energy is about $5\%$.  Extrapolating  data to the energy  $\sqrt{s}=13$ TeV one can get an estimate for the bound on $\sigma_{diff}(s)$ and the reflective scattering  contribution to it  at the level of $28.2$ mb and  $(6-8)\%$, respectively\footnote{The extrapolated value of $r(s)$ at this energy is about $0.3$ fm.}.  It would be interesting to confront these estimates to the experimental data for
 $\sigma_{diff}(s)$ at the above energy. The data at  $\sqrt{s}=7$ TeV \cite{lipari} are in agreement with this bound.
 It is useful to consider an inverted energy evolution, i.e. the case of the decreasing energy, also. Due to the supposed monotonous energy dependence, the function $r(s)$ will be moving to zero with the energy decrease and it can only be approximated by this zero  value at lower energies ($s<s_r$) because the negative values of the impact parameter have no sense (cf. Fig. 4).
\begin{figure}[hbt]
\begin{center}
\resizebox{7cm}{!}{\includegraphics*{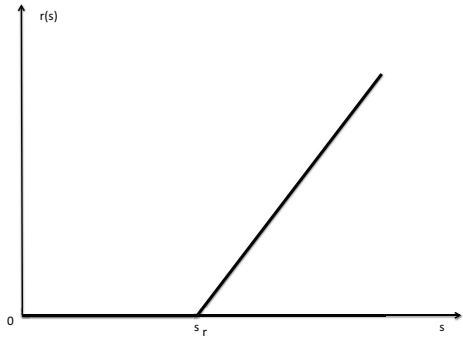}}
\end{center}
\vspace{-1cm}
\caption[ch1]{\small Schematic energy dependence of the function  $b=r(s)$.}
\end{figure} 
 Thus, the function 
  $2\pi\int_{r(s)}^\infty bdb[1-S(s,b)]$
  will be transformed into 
 $4\pi\int_{0}^\infty bdbf(s,b)$
 at $s\leq s_r$, (note that $1-S=2f$). The latter  is just $\sigma_{tot}(s)/2$ and, therefore, the bound Eq.(\ref{brfd}) is being converted into the standard  Pumplin bound. 
 This demonstrates selfconsistency of the above considerations and universality of Eq. (\ref{brfd}), i.e. it is valid at low and high energies.
 
 Thus, the Eq.(\ref{brfd}) should be considered as a generalization of the  Pumplin bound for the inelastic diffraction cross-section.  Eq.(\ref{brfd}) is valid 
 in the whole range of the elastic amplitude variation allowed by unitarity and, in particular, in the energy region where the black-disc limit is exceeded. The above generalization has a practical meaning since there is a definite indication that the black-disc limit has been exceeded in the central hadron collisions at the LHC energy $\sqrt{s}=7$ TeV \cite{alkin}. Despite that the asymptotic regime is not reached at the LHC energies, there are certain indications on the turn-on of the transitional regime at these energies where elastic scattering starts to dominate at small impact parameters. Thus, the measurements at $\sqrt{s}=13$ TeV are in the energy range where the generalized upper bound for the inelastic diffractive cross-section should be applied. The reflective scattering mode could also be responsible for the interesting phenomena observed uder cosmic ray studies \cite{crays}. 
 \section{The model estimates} 
 The unitary model for the  $S(s,b)$ can also be used  to estimate qualitatively the dependencies of the cross-sections 
 $ \sigma_{diff}(s)$ and 
 $\sigma_{ndiff}(s)$. The reflective scattering is a characteristic picture of the  model presented below. 
 It is based on the rational form of the unitarization and represents the function $S(s,b)$ 
 in the form:
 \begin{equation}\label{um}
 S(s,b)=   {[1-U(s,b)]}/{[1+U(s,b)]},
  \end{equation}   
 The $U(s,b)$ is the generalized reaction matrix element, which is considered to be an
input dynamical quantity and it is taken to be a real function.  
The various dynamical models can be used for the function $U(s,b)$. To get the qualitative estimates we are using
 the simplified form of this function
which conforms to rising total cross-section and analytical properties over the transferred momentum, i.e.
  \begin{equation}\label{umf}
 U(s,b) = g(s) \exp({-\mu b}),
  \end{equation}  
  where $g(s)\sim  s^\lambda $ , $\lambda$  and $\mu$ are some constants. 
  Eq. (\ref{umf}) resembles form used   by Heisenberg in his model for the total cross--sections of the inelastic processes \cite{heis}. 
  However, the model is relevant for the black-disc limit saturation only
  (cf. e.g. \cite{dosch} and references therein) and it does not include  
  elastic scattering and effects of self-damping of the inelastic channels \cite{bak}. 
  
  Then the following asymptotical dependencies will take place\footnote{The explicit expressions for $r(s)$ and $\sigma_{inel}(s)$  are the following    
 \[r(s)=\frac{1}{\mu}\ln g(s)\,\,\mbox{and}\,\, \sigma_{inel}(s)=\frac{8\pi}{\mu^2}\ln(1+g(s)). \]}:
   \begin{equation}\label{asym}
 \sigma_{tot}(s) \sim \ln^2 s, \, \, \sigma_{el}(s) \sim \ln^2 s,  \, \, \sigma_{inel}(s) \sim \ln s  \, \,    \mbox{and} \, \, r(s) \sim \ln s .
   \end{equation}  
   From Eq.  ( \ref{brfd}) it follows that for the ratio ${\sigma_{diff}(s) }/{\sigma_{inel}(s)}$ the inequality takes place  
   \begin{equation}\label{brfd1}
\frac {\sigma_{diff}(s) }{\sigma_{inel}(s)}\leq 1- \frac{2\pi}{\sigma_{inel}(s)}\int_{r(s)}^\infty bdb[1-S(s,b)].
 \end{equation}
From Eqs.  ( \ref{brfnd}) and  ( \ref{asym}) it follows that    $ \sigma_{ndiff}(s) \sim \ln s  $ and  the second term in Eq. (\ref{brfd1})
tends to $1/2$ at $s\to \infty$.  Thus, in this approach both parts of $ \sigma_{inel}(s)$ would have similar asymptotical energy dependencies, which are proportional  to   $ \ln s  $.

Eq. (\ref{brfd1}) can be further simplified if one notes that at large values of $b$ 
\[
1-S(s,b)\simeq 2h_{inel}(s,b), 
\]
 $h_{inel}(s,b)$ has its maximal value at $b=r(s)$ and \[\sigma_{inel}\simeq \frac{8\pi}{\mu^2}\ln g(s)\]
 at $s\to\infty$: 
  \begin{equation}\label{brfd2}
\frac {\sigma_{diff}(s) }{\sigma_{inel}(s)}\leq 1- \frac{\mu}{2}\int_{r(s)}^\infty dbh_{inel}(s,b).
 \end{equation} 
 The limiting value of the inelastic overlap function integral over $b$ \[\int_{r(s)}^\infty dbh_{inel}(s,b)\] in the model is $1/\mu$ and bound on
 inelastic diffractive cross takes the simplest form
 
  \begin{equation}\label{brfd3}
\frac {\sigma_{diff}(s) }{\sigma_{inel}(s)}\leq 1/2.
 \end{equation} 
 Finally, one can assume  saturation of the bound Eq. (\ref{brfd3}) and arrive that way to the asymptotic equipartition of the inelastic cross--section into diffractive and non-diffractive parts. 
 
\section*{Conclusion}
The generalized upper bound Eq.(\ref{brfd}) for the inelastic diffraction has been derived. It has also been shown  that
there is no  inconsistency between saturation of the unitarity limit leading to Eq. (\ref{rd}) and this  bound on the inelastic
diffractive cross--section. The obtained energy-independent ratio 
$\sigma_{diff}(s)/\sigma_{inel}(s)$ conforms to the definition of the inelastic diffraction as a result of the 
Pomeron exchanges 
as well as to the recent experimental trends observed at the LHC.  
This allows one to reconcile the 
results of $s$- and $t$-channel approaches to the inelastic diffraction cross-section.

If one assumes  any mechanism resulting in saturation of the black-disc limit at the asymptotic energies, it is difficult to ensure  such reconcilement   already in the LHC energy range.
The  LHC experiments
at the new higher  energies would be definitely helpful  for resolving  picture of the inelastic diffraction and elastic scattering at $s\to\infty$.

\end{document}